# To jump or not to jump: The Bereitschaftspotential required to jump into 192-meter abyss

Nann, M.[1], Cohen L.G.[2], Deecke, L.[3], Soekadar, S.R.[1*]

[1] Applied Neurotechnology Laboratory, Department of Psychiatry and Psychotherapy, University Hospital of Tübingen, Germany
[2] National Institute of Neurological Disorders and Stroke (NINDS), Human Cortical Physiology and Stroke Neurorehabilitation Section, Bethesda, Maryland, USA
[3] Department of Clinical Neurology, Medical University Vienna, Austria

[*]Correspondence to: Surjo Soekadar, MD, Applied Neurotechnology Laboratory, Department of Psychiatry and Psychotherapy, University Hospital of Tübingen, Calwerstr. 14, 72076 Tübingen, Germany. Tel: +49 7071 2982640; E-mail: surjo@soekadar.com

**Manuscript statistics:**

| | |
|---|---|
| Number of figures: | 2 |
| Number of pages: | 11 |
| Number of words: | Title: 14 |
| | Abstract: 143 |
| | Main text: 1869 |
| | Methods: 555 |
| Number of references: | 28 |

**Abstract**

Self-initiated voluntary acts, such as pressing a button, are preceded by a negative electrical brain potential, the Bereitschaftspotential (BP), that can be recorded over the human scalp using electroencephalography (EEG). Up to now, the BP required to initiate voluntary acts has only been recorded under well-controlled laboratory conditions.

It is thus not known if this form of brain activity also underlies motor initiation in possible life-threatening decision making, such as jumping into a 192-meter abyss, an act requiring extraordinary willpower. Here, we report BP before self-initiated 192-meter extreme bungee jumping across two semi-professional cliff divers (both male, mean age 19.3 years). We found that the spatiotemporal dynamics of the BP is comparable to that recorded under laboratory conditions. These results, possible through recent advancements in wireless and portable EEG technology, document for the first time pre-movement brain activity preceding possible life-threatening decision making.

**Introduction**

The decision to perform a self-initiated act, such as releasing an arrow during Kyūdō (Japanese art of archery), pushing off for the first glide in alpine (downhill) skiing or jumping from a rock in cliff diving, requires integration and synchronization of internal and external sensory information[1], as well as a sense for the opportune moment for action (a moment termed καιρός in Ancient Greek, e.g. Ilias IV, 185)[2]. Developing this sense often requires many years of training, particularly when motor initiation involves possible life-threatening decision making.

For a long time, the neural origins of self-initiated acts remained an enigma. While neurophysiological experimentations up to the 1960ies were mainly influenced by behaviourism focusing on stimulus-response paradigms, the discovery of the Bereitschaftspotential (BP) in 1965 (engl. readiness potential)[3,4] signified an entirely new direction in neurophysiological research investigating the neural substrate of *self-initiated voluntary* acts, i.e. acts that are not triggered by external stimuli.

Due to the non-stationary and dynamic nature of brain activity, stimulus-response paradigms usually require averaging of brain activity over multiple trials. This can be easily achieved by using the stimulus as the common starting point (trigger) for averaging. Due to lack of a known starting point, however, such approach cannot be applied to unpredictable self-initiated movements. Building on early EEG systems that used ferromagnetic tapes to store the recorded signals, an important technological innovation and prerequisite for the discovery of the BP was to reverse such tapes and use the onset of electromyographic (EMG) activity related to a volitional act, e.g. moving a finger, as marker for reverse computation ("Rückwärtsanalyse") of the EEG signal[5]. In the original experiment, EEG signals of approximately 250 trials were averaged and evidenced a negative electric potential building up approximately 1.5 s before the onset of EMG activity (Fig. 1A). Study participants had to sit still in a Faraday cage with their head reclined into a headrest to avoid EEG artefacts (Fig. 1B). The recorded brain potential was most evident at the vertex electrode and reached amplitudes of up to 10-15 µV[3,4]. Based on the waveform and involved generators, two phases of the BP can be distinguished: the "early BP" recordable over the supplementary motor area (SMA), pre-SMA and lateral premotor cortex beginning about 1.5 s before movement onset, and the "late BP" with a steep negative slope recordable over the contralateral primary motor cortex (M1) and lateral premotor cortex beginning approximately 400 ms before movement onset[6,7].

Using a multimodal neuroimaging approach based on functional magnet resonance imaging (fMRI) and electroencephalography (EEG), Cunnigton et al.[8] showed a specific correlation between global electric field power of pre-movement neural activity and the metabolic (i.e. blood-oxygen-level dependent, BOLD) activity of the anterior mid-cingulate cortex (aMCC). By applying dynamic causal modelling, they found strong reciprocal interactions between the aMCC and the supplementary motor areas (SMA) that they identified to be important for the sustained activity during the early BP. Moreover, besides involvement of the cortico-basal ganglia-thalamo-cortical loop (motor loop), a number of studies indicate that the prefrontal cortex (PFC), a region that was found to be tightly linked to the concept of willpower[9] (i.e. the capacity to override an unwanted thought, feeling or impulse), decision making and executive control[10], plays a critical role for initiation of voluntary movements[11]. Particularly the medial part of the PFC (mPFC) was shown to be important for regulating neuronal circuits linked to fear and anxiety[12].

While it was shown that the magnitude and waveform of the BP depend on various factors, such as force exerted, speed and precision of movement as well as pace of movement repetitions or complexity of movement[13], it was unknown whether the BP also underlies movement initiation in possible life-threatening decision making, such as self-initiating a jump into a 192-meter abyss. The main reason for this gap in knowledge is that, despite numerous replications and several thousand publications since its discovery, the BP has never been recorded in a real-life situation outside the laboratory where motor initiation in possible life-threatening decision making can be investigated.

Today, more than 50 years after the discovery of the BP, EEG technology has improved across multiple domains, particularly in terms of miniaturization, digitization and wireless signal transmission. State-of-the-art wireless EEG systems can record from up to 64-channels with 24-bit A/D conversion and input noise peak-to-peak levels below 2 $\mu$V. Especially, the development of active EEG electrodes, i.e. electrodes with build-in readout circuitry that locally amplify and buffer EEG signals before transmitting the signals through cablings, have improved signal quality[14]. These advancements fostered the development of innovative neurotechnologies, such as brain-machine interfaces (BMIs) that translate brain activity into control commands of external machines, robots or computers. Recently, it was shown, for example, that severely paralyzed quadriplegics can operate an EEG-controlled brain/neural hand-exoskeleton (B/NHE) to perform activities of daily living, such as eating and drinking in an outside restaurant[15]. We, thus, reasoned that, under optimal conditions, recording of the BP outside the laboratory might be feasible, even in extreme real-life scenarios such as 192-meter bungee jumping.

Deriving from an ancient ritual on Vanuatu, an island in the South Pacific Ocean, bungee jumping, i.e. jumping from a tall structure while fixed to an elastic cord, has become a popular and commercialized activity with millions of jumps since the 1980ies. Besides representing a test of courage (bungee jumping was seen as an expression of boldness in ancient Vanuatu), it was shown that bungee jumping can result in a marked increase of euphoria ratings and concentration of beta-endorphin (by more than 200 %) measured immediately after a bungee jump[16]. Here, we report successful recording of the BP across two semi-professional cliff divers who performed several bungee jumps from a 192-meter bungee platform (the second highest bungee platform in Europe). To evaluate the impact of possible life-threatening decision making on the BP's spatiotemporal dynamics, such as onset and waveform, BPs recorded before bungee jumping were compared to BPs recorded before the same subjects jumped from a 1-meter block.

**Results**

Analysis of EEG recordings showed a clear negative deflection beginning approximately 1.5 s before movement onset (Fig. 2A). The maximum EEG deflection across all bungee jumps ranged at 17.62±1.09 $\mu$V.

Further analysis evidenced successful detection of a BP before bungee jumping (jumper 1: $M$ = -19.23 $\mu$V, $SD$ = 9.21 $\mu$V, $t(9)$ = -2.087, $p$ = .033; jumper 2: $M$ = -10.49 $\mu$V, $SD$ = 4.12 $\mu$V, $t(11)$ = -2.548, $p$ = .014) and jumping from a 1-meter block (jumper 1: $M$ = -11.63 $\mu$V, $SD$ =

2.71 µV, *t*(14) = -4.293, *p* < .001; jumper 2: *M* = -8.37 µV, *SD* = 3.86 µV, *t*(11) = -2.167, *p* = .027).

We found no difference in BP onset comparing BPs recorded before bungee jumping (jumper 1: *Mdn* = -1.81 s; jumper 2: *Mdn* = -1.29 s) and BPs recorded before jumping from a 1-meter block (jumper 1: *Mdn* = -1.80 s; jumper 2: *Mdn* = -1.06 s) (jumper 1: *U* = 75, *p* = .978, jumper 2: *U* = 72, *p* = .684).

Moreover, comparison between BP waveforms recorded before bungee jumping and BP waveforms recorded before 1-meter block jumping showed a strong correlation across both jumpers (jumper 1: $r_s$ = .639, *p* < .001; jumper 2: $r_s$ = .80, *p* < .001).

**Discussion**

To our knowledge, this is the first report on successful recordings of the BP outside the laboratory and in the context of self-initiating a voluntary act in possible life-threatening decision making. We found that the BP can be successfully detected by averaging EEG data from less than 15 trials documenting feasibility of such recordings outside the laboratory. BPs recorded before self-initiated 192-meter bungee jumps involving possibly life-threatening decision making and BPs recorded before 1-meter block jumps under laboratory conditions showed comparable spatiotemporal dynamics. While the necessarily low number of trials demonstrates that the pre-bungee jumping BP is very well expressed, the small number of subjects and trials constraints the possibility to generalize these findings towards other real-life scenarios.

Catalysed by Benjamin Libet's experiments in the 1980ies[17] indicating that the conscious decision for a self-initiated act occurs not earlier than 200-250 ms before the onset of EMG activity, the discovery of the BP led to a vivid and controversial discussion about the free will[18] (it should be noted that neither Kornhuber and Deecke[19] nor Libet ever called human freedom or free will into question[20]. While Kornhuber and Deecke admit that "absolute freedom from nature is an impossibility"[19] leaving humans relative freedom, or freedom in degrees, Libet argued that, even if a volitional process is initiated unconsciously, the conscious function can still control the outcome by exerting a *veto*). While it is unlikely that neurophysiological experiments can decisively contribute to solving this controversy, feasibility of BP recordings in real-life scenarios allow for investigating the neural mechanisms underlying self-initiated voluntary acts under realistic conditions. In this context, EEG recordings using a larger number of electrodes (≥ 32) and application of advanced signal processing techniques, such as connectivity[11] or entropy measures[21,22], may provide important insights regarding the involved brain networks and their interactions. Moreover, implementation of real-time signal processing within the framework of brain-computer interfaces (BCIs) allows answering some profound research questions, for example determining the point in time at which people are still able to cancel ("veto") a movement after the elicitation of a BP[23,24]. Moreover, establishing BCIs based on motor-related cortical potentials (MRCP)[25] that include the BP to detect movement intentions in everyday life environments may critically enhance existing BCI systems, e.g. to restore activities of daily living to quadriplegics[15] or in the context of neurorehabilitation[26]. Besides further substantiating that neurotechnologies, i.e. technological tools to interact with the brain, are now about to enter everyday life environments[27], our report shows that brain physiological assessments previously regarded unfeasible, e.g. assessment of the BP outside the laboratory and under extreme conditions, are viable and can extend the possibility to test scientific hypotheses that were previously un-testable.

It could be argued that, in most lab-based investigations, decision-making, i.e. the decision to obey the instruction to self-initiate a particular movement, e.g. to press a button, was already concluded before the beginning of the actual experiment. If so, subjects have only to decide *when* to press and not *whether* to press. In bungee jumping or other extreme activities that involve possibly life-threatening decision making, this decision-making process cannot be assumed to be concluded at any point in time before the actual jump. Many bungee jumping novices interrupt their first attempt and need an external trigger to overrule their inner resistance. Even bungee jumpers with year-long experience and dozens of jumps report that each jump requires them to overcome this inner resistance suggesting a decisive role of the PFC coordinating neuronal circuits related to willpower, executive functions and fear responses. While our data indicate that this coordination process does not influence the onset or waveform of the pre-bungee jumping BP, the limited number of electrodes and trials did not allow for analyses of fronto-parietal cortico-cortical interactions. Further studies are needed to elucidate these possible network interactions and their relationship to the generation of the BP. In this context, use of virtual reality may be particularly helpful as a complementary approach to investigate the neural origins of self-initiated acts also in laboratory environments.

**Methods**

*Experimental setup*

Two semi-professional cliff divers (both males, mean age 19.3 years) performed self-initiated voluntary jumps, once outside the laboratory from a 192-meter bungee jumping platform (operator: Rupert Hirner Bungee Jumping GmbH, Europa Bridge in Innsbruck, Austria; http://www.europabruecke.at/), and once under well-controlled laboratory conditions from a 1-meter block. The jumpers were informed by the platform operator that they were free to jump as often as they liked, and that all jumps were entirely voluntary and could be aborted at any time.

Before the jumps, specific instructions on how to perform the jumps without generating excessive muscle artefacts and a clear trigger signal for reverse computation were provided (e.g. keeping head motions and blinking to a minimum, relaxing the arms and trunk, initiating the jump by coming up on the toes and bending forward). Besides allowing for precise detection of movement onset due to an initial upward movement, muscle artefacts were kept at a minimum by initiating the jump with the toe extensors (the most distant muscles from the EEG recording sites).

*Portable electroencephalography (EEG) and signal processing*

For EEG recordings, a wireless and portable 8-channel EEG system (LiveAmp®, Brain Products GmbH, Gilching, Germany) in combination with an active electrode system (actiCAP®, Brain Products GmbH) was used. EEG was recorded from eight conventional recording sites (Fz, FC1, FC2, C3, Cz, C4, CP1, CP2 according to international 10/20 system). Ground and reference electrodes were placed at Fpz and the mastoid, respectively. Impedance of all electrodes was kept below 5 kΩ to ensure high EEG signal quality. For precise detection of movement onset, a three-axis acceleration sensor (± 2 g) was used. The EEG amplifier and acceleration sensor [dimensions (w x d x h): 83 x 51 x 14 mm, overall weight: approx. 60 g (incl. built-in battery)] were fixed to the jumper's occiput using customized electrode caps (EasyCap®, Herrsching, Germany), adhesive tape and an elastic net dressing. Besides allowing for EEG recordings for up to 3 hours, the system transmitted the recorded signals wirelessly via Bluetooth to a signal processing unit (Sony Vaio Duo 13® equipped with an Intel Core i7® processor). EEG and accelerometer data were sampled at 250 Hz and band-pass-filtered at 0.1 to 3 and 5 Hz, respectively. EEG signals recorded from Cz before each jump (defined as trial) were time locked to the onset of an accelerometer signal exceeding the 95 % confidence interval recorded during movement preparation at rest (defined as detected movement onset) (Fig. 2B) and then epoched into 3-second windows. Epochs ranged from -2.5 s to +0.5 s with a baseline correction relative to the first 0.5 s (reference window). Epochs with non-physiological signal amplitudes exceeding ±100 µV or large drifts before detected movement onset were excluded from further analysis.

*Outcome measures and statistics*

Successful detection of a BP was defined as negative deflection of the EEG signal 400 ms before movement onset across all trials[7]. The BP onset was defined as time point the averaged

EEG signal evidenced a continuous negative deflection for more than 500 ms. Statistical differences between BP onsets were assessed using the Mann-Whitney-U test. Similarity of waveform was defined as a Spearman's rank correlation coefficient of higher than 0.6 (i.e. strong correlation) when comparing BPs recorded before bungee jumping and 1-meter block jumping. Significance level was set to $p$ = .05 for all analyses.

**Acknowledgements**

We thank Manuel Halbisch and Pascal Pollin for their willingness to have their brain activity recorded before their bungee jumps, and Patrick Britz, Luke Enge, Liam Scannell and Thomas Emmerling (Brain Products GmbH, Gilching, Germany) for their support in preparing and conducting the reported physiological recordings. S.R.S. received special support by the Brain & Behavior Research Foundation as 2017 NARSAD Young Investigator Grant recipient and P&S Fund Investigator, and M.N. was supported by the Baden-Württemberg Stiftung (NEU007/1).

**Author contribution**

M.N., L.D. and S.R.S. collected the data. M.N. analysed the data. M.N., L.D. and S.R.S. interpreted the data, performed the literature search and wrote the manuscript. M.N. and S.R.S created the figures. M.N., L.G.C., L.D. and S.R.S. edited the manuscript.

**Competing financial interests**

None declared.

**Figure legends**

**Fig. 1A:** Reverse-computation ("Rückwärtsanalyse") of electroencephalographic (EEG) signals before self-initiated voluntary finger movements (right index finger flexion) evidenced a negative potential shift beginning approximately 1.5 s before finger movement-related electromyographic (EMG) activity could be detected. This characteristic EEG signal was termed "Bereitschaftspotential" (BP) (in engl. also readiness potential) by Kornhuber and Deecke [3] and provided first important insights to the neural origins of self-initiated acts[3,4] (superposition of 6 experiments of the same subject at 6 different days with 250 trials each, i.e. 1500 self-initiated finger movements (right index finger flexion) of the same subject B.L., adapted from Deecke, et al. [28]). **1B:** Subjects had to sit still in a Faraday cage with the head reclined into a headrest to avoid EEG artefact (photograph from the original experimental setup used in Kornhuber and Deecke [3], provided by Lüder Deecke, Austria).

**Fig. 2A.** Electroencephalographic (EEG) recordings before 192-meter bungee jumping evidenced a negative potential shift over the vertex electrode (Cz) with the characteristic features of the Bereitschaftspotential (BP) (orange: jumper 1, blue: jumper 2; average over all trials). The 95 % confidence interval is illustrated as orange and blue shaded areas, respectively. **2B:** Movement onset before bungee jumping was detected by an accelerometer integrated into the EEG system. The solid line shows the averaged accelerometer signal across all trials of both jumpers. The 95 % confidence interval is indicted by the shaded areas. Reverse-computation of the pre-bungee jumping BPs was time-locked to the detected movement onset. **2C:** One of the semi-professional cliff divers in pre-bungee jumping posture. 192-meter bungee jumps were initiated by coming up on the toes and bending forward. The EEG system was attached to the jumper's occiput using a customized electrode cap (EasyCap®, Herrsching, Germany), adhesive tape and an elastic net dressing.

**Figures**

**Figure 1.**

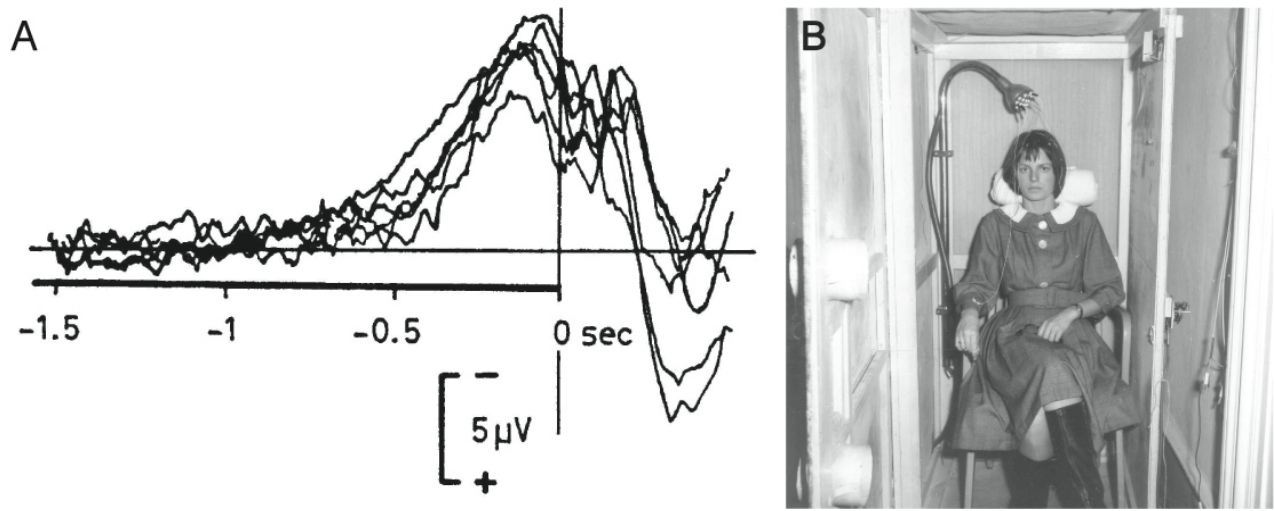

**Figure 2.**

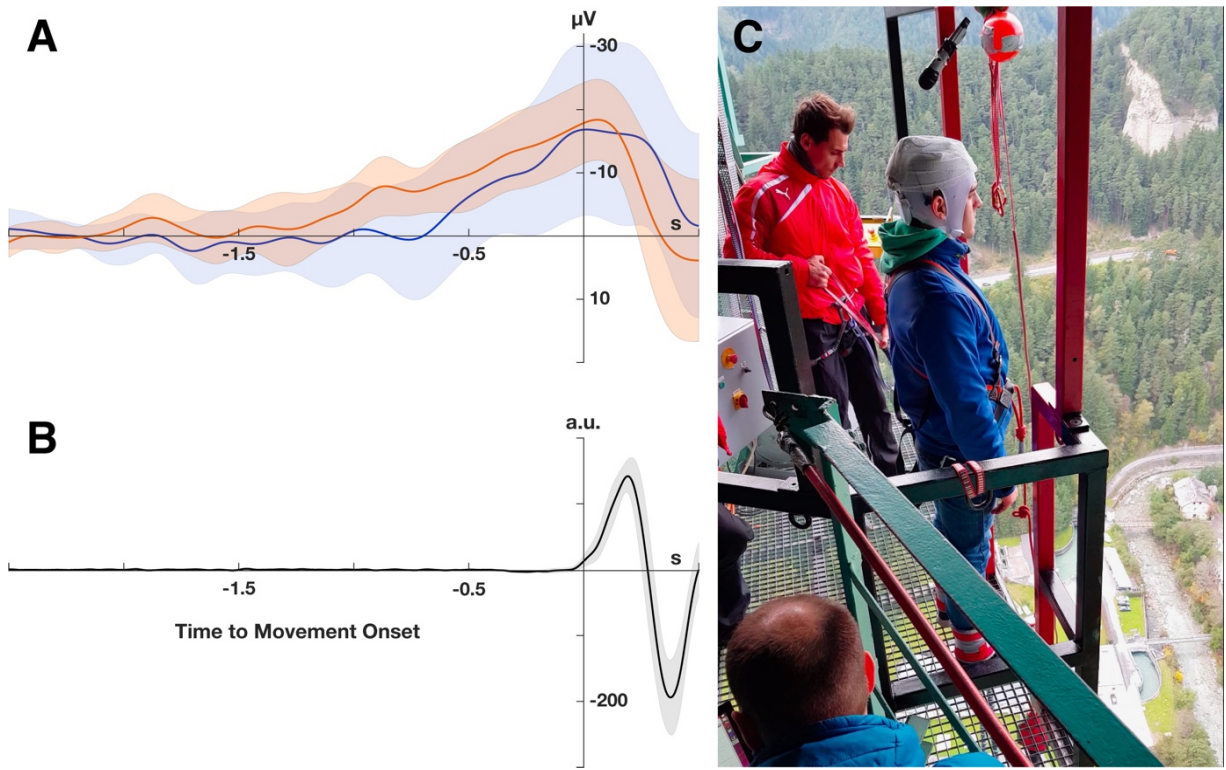

Time to Movement Onset